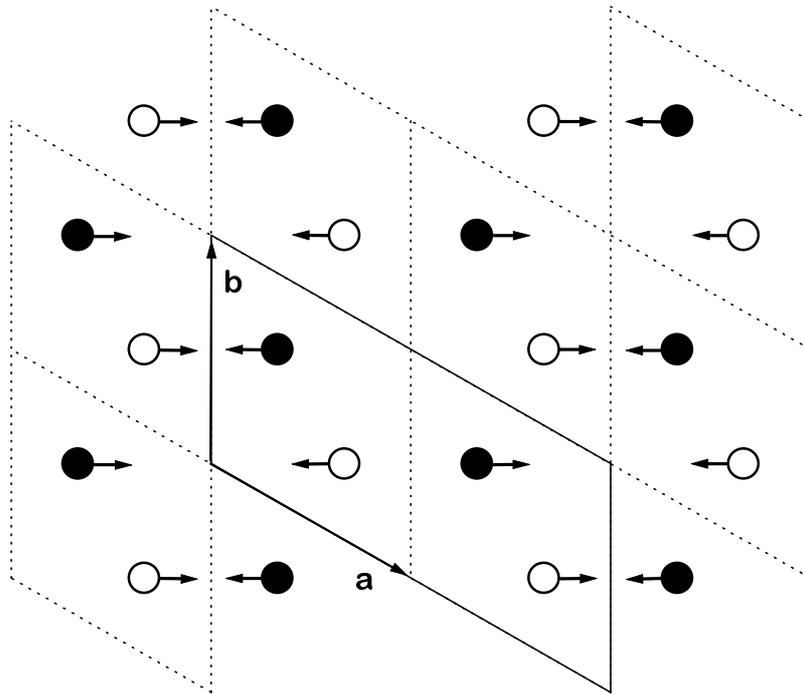

Fig. 1

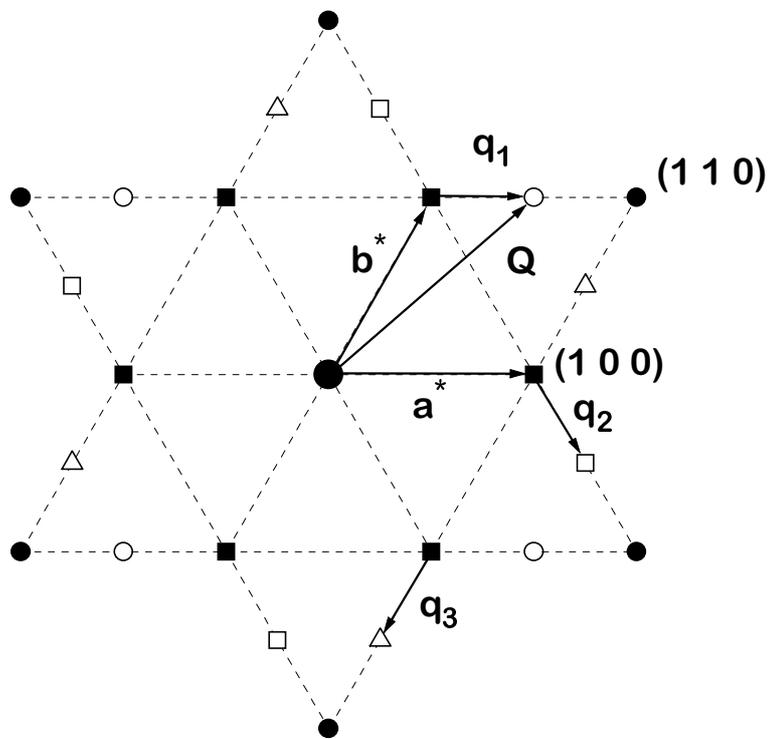

Fig. 2

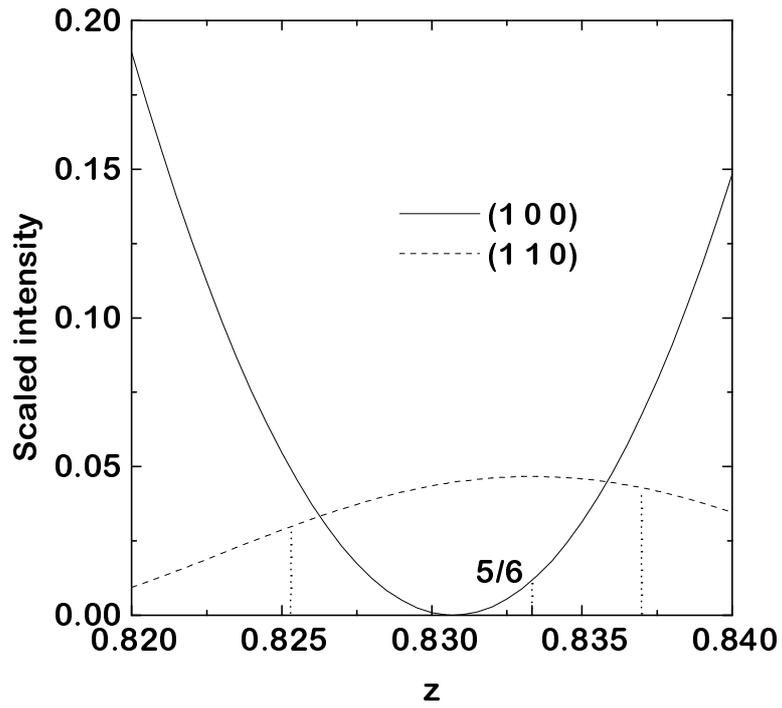

Fig. 3

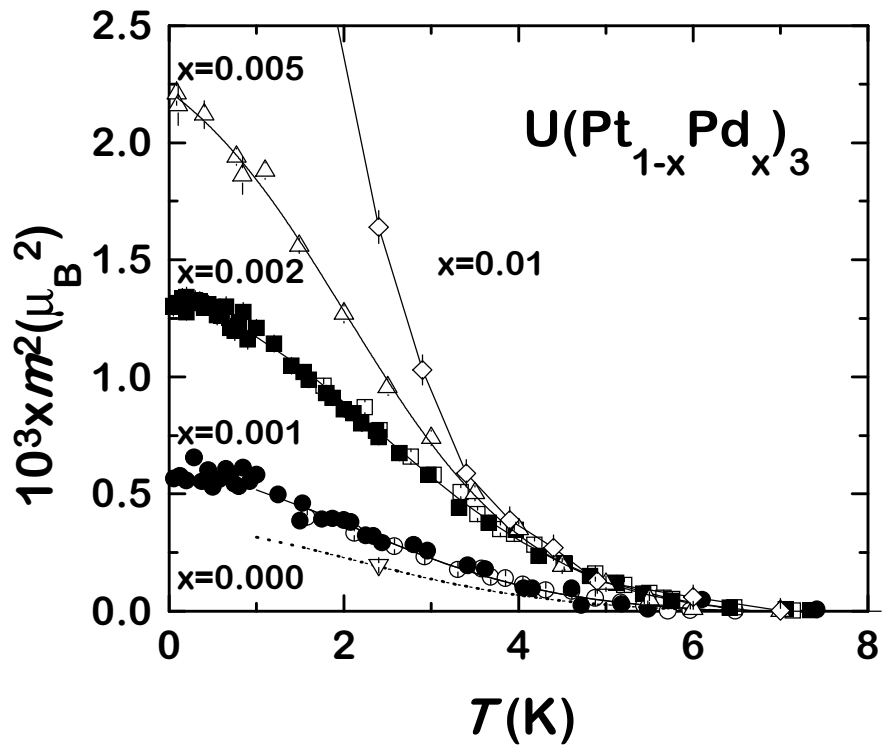

Fig. 4

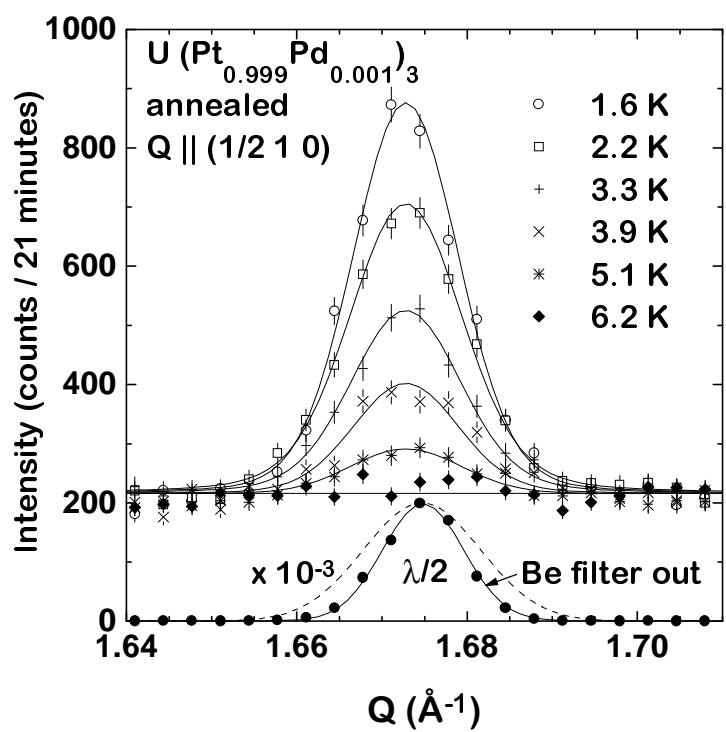

Fig. 5

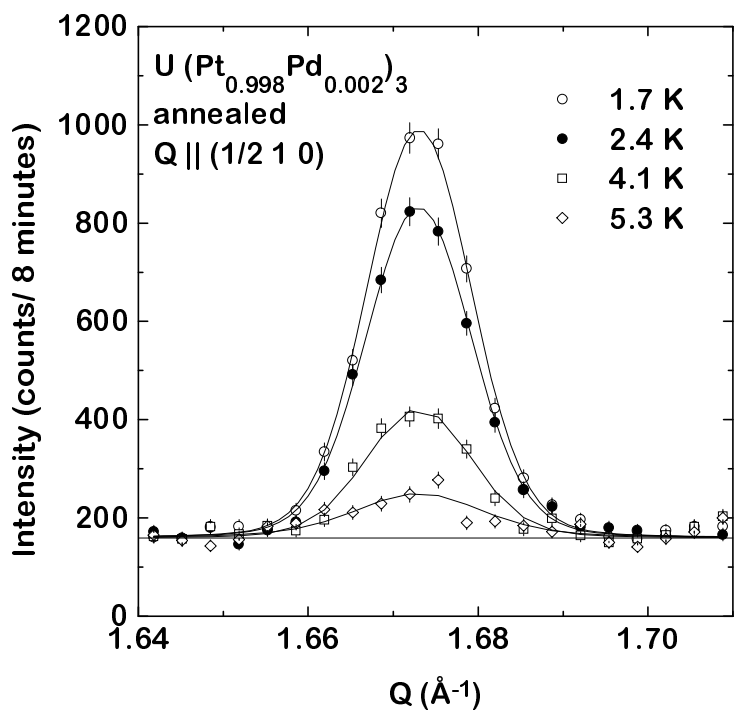

Fig. 6

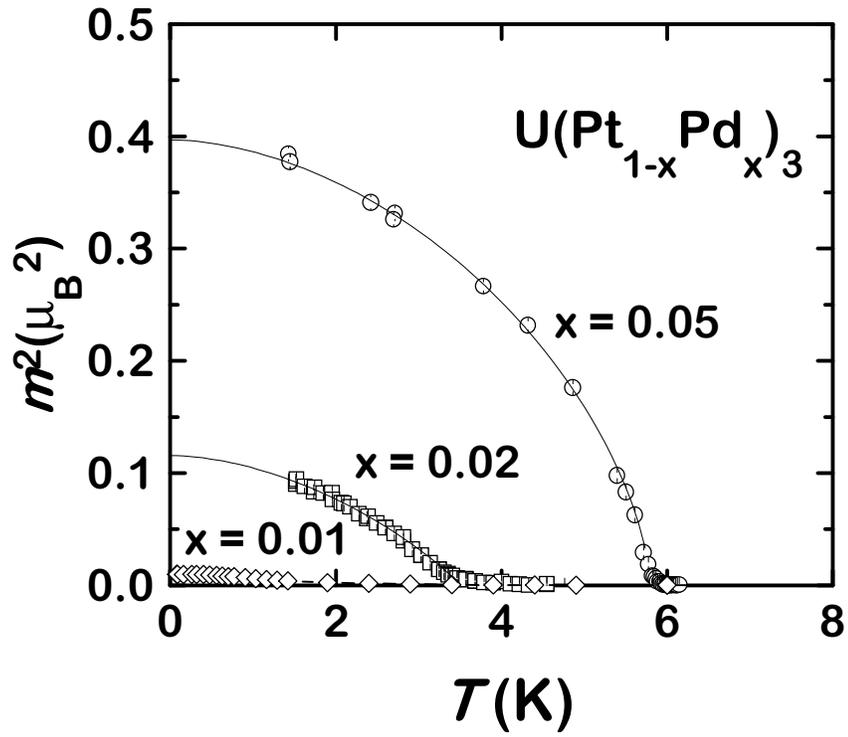

Fig. 7

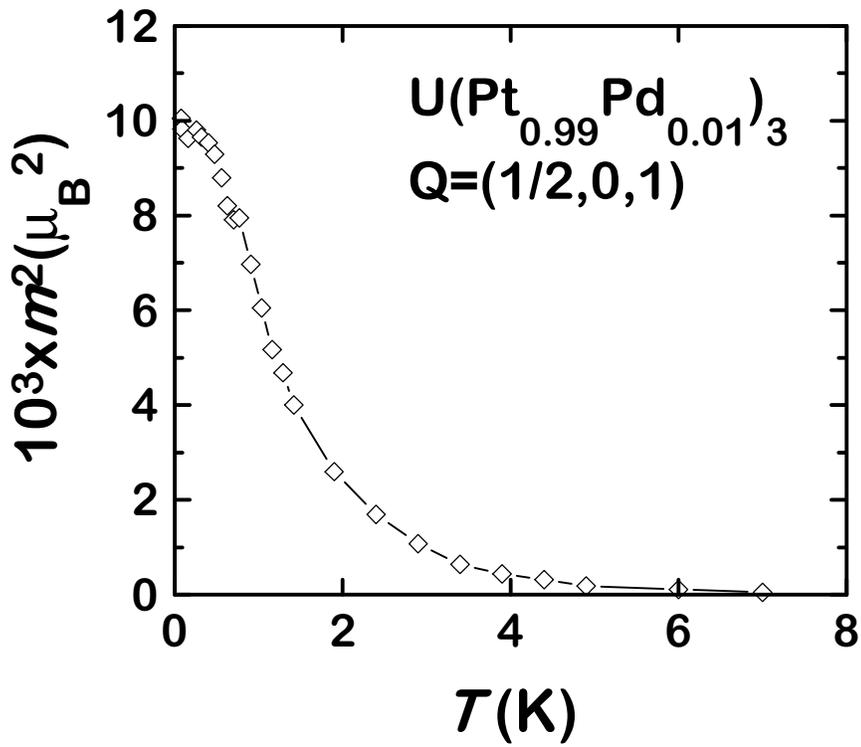

Fig. 8

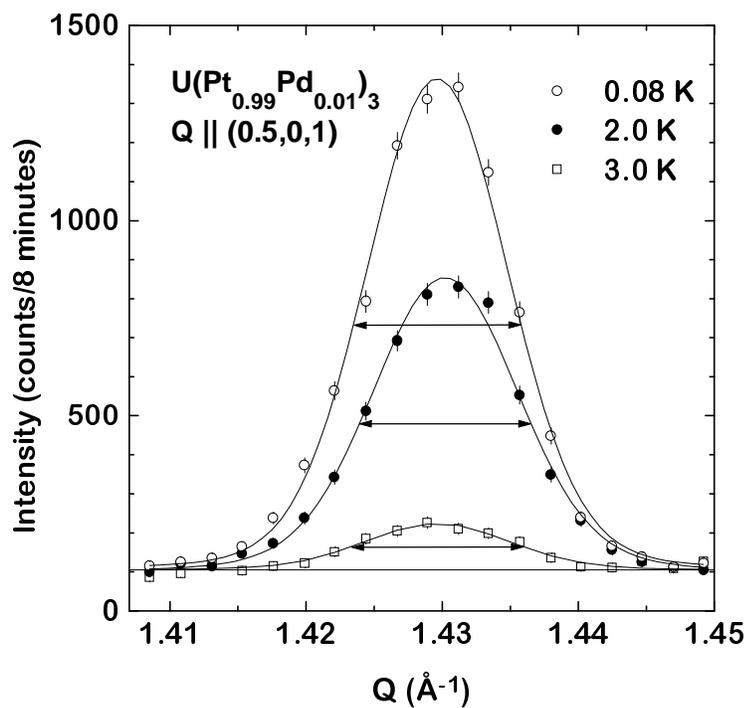

Fig. 9

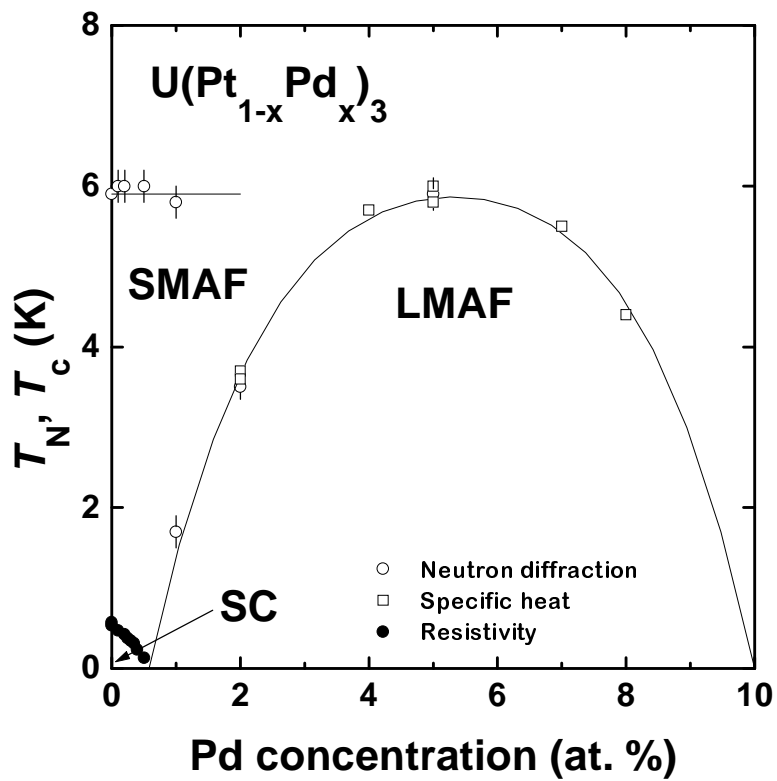

Fig. 10

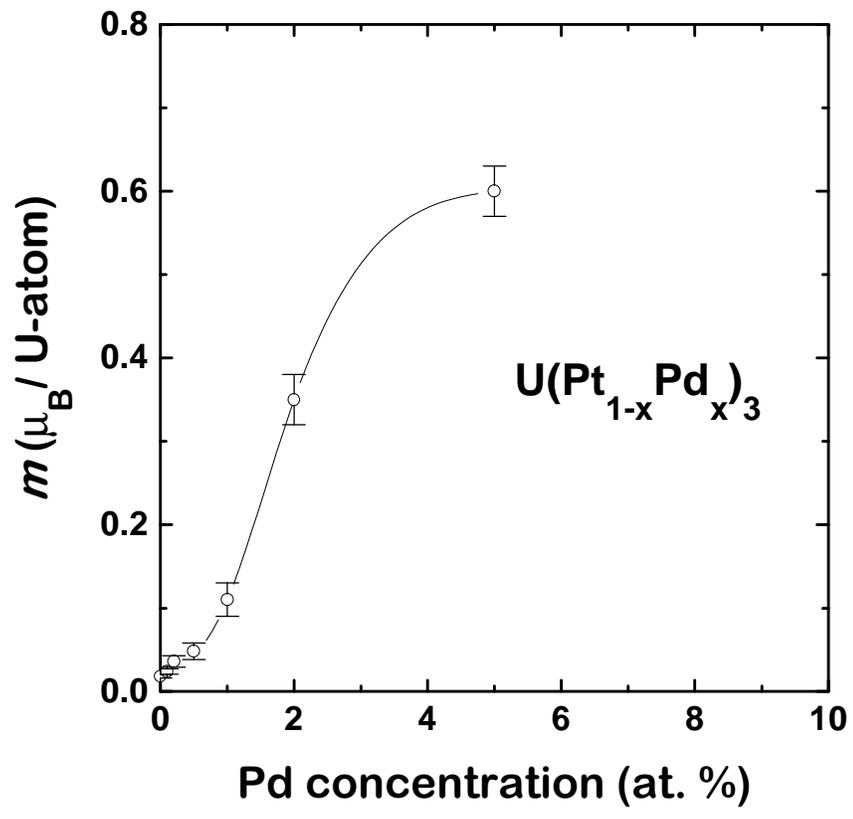

Fig. 11

# Neutron-diffraction study of the evolution of antiferromagnetic order in UPt$_3$ doped with Pd

R.J. Keizer, A. de Visser, A.A. Menovsky and J.J.M. Franse
*Van der Waals-Zeeman Institute, University of Amsterdam,*
*Valckenierstraat 65, 1018 XE Amsterdam, The Netherlands*

B. Fåk
*Département de Recherche sur la Matière Condensée, SPSMS/MDN*
*CEA, 38054 Grenoble, France*

J.-M. Mignot
*Laboratoire Léon Brillouin, CEA-CNRS, CEA/ Saclay, 91191 Gif sur Yvette, France*

Neutron-diffraction experiments have been carried out on a series of heavy-electron pseudobinary U(Pt$_{1-x}$Pd$_x$)$_3$ single crystals ($x \leq 0.05$). The small-moment antiferromagnetic order reported for pure UPt$_3$ is robust upon doping with Pd and persists till at least $x = 0.005$. The ordered moment grows from 0.018±0.002 $\mu_B$/U-atom for pure UPt$_3$ to 0.048±0.008 $\mu_B$/U-atom for $x = 0.005$. The Néel temperature, $T_N$, is approximately 6 K and, most remarkably, does not vary with Pd contents. The order parameter for the small-moment antiferromagnetism has an unusual quasi-linear temperature variation. For $x \geq 0.01$ a second antiferromagnetic phase with much larger ordered moments is found. For this phase at optimum doping ($x = 0.05$) $T_N$ attains a maximum value of 5.8 K and the ordered moment equals 0.63±0.05 $\mu_B$/U-atom. $T_N(x)$ for the large-moment antiferromagnetic order follows a Doniach-type phase diagram. From this diagram we infer that the antiferromagnetic instability in U(Pt$_{1-x}$Pd$_x$)$_3$ is located in the range 0.5-1 at.% Pd.

Corresponding author:   R.J. Keizer
                        Van der Waals-Zeeman Institute
                        University of Amsterdam
                        Valckenierstraat 65
                        1018 XE Amsterdam
                        The Netherlands
                        Phone: 31-20-5255795
                        Fax:   31-20-5255788
                        E-mail: rjkeizer@phys.uva.nl

# 1. Introduction

It has been recognized, for more than a decade now, that the heavy-electron compound $UPt_3$ is close to an antiferromagnetic instability. Evidence for the proximity to a magnetic instability is provided by pronounced spin-fluctuation phenomena at low temperatures [1] and incipient magnetic ordering [2], which can readily be made visible by chemical substitution. The low-temperature thermal, magnetic and transport properties of pure $UPt_3$ demonstrate the formation of a strongly renormalized Fermi liquid at low temperatures [1-3]. The coefficient, $\gamma= 0.42$ J/molK$^2$, of the linear term in the specific heat, $c(T)$, is very much enhanced with respect to a normal metal, which gives rise to a Fermi-liquid description with a quasiparticle mass of ~200 times the free electron mass. The low-temperature Pauli susceptibility, $\chi_0=\chi(T\rightarrow 0)$, is equally enhanced. Upon raising the temperature, $\chi(T)$ exhibits a maximum at $T_{max}= 18$ K, which indicates the stabilization of antiferromagnetic spin fluctuations below $T_{max}$. From the electrical resistivity, $\rho(T)$, data, it follows that the coherence regime sets in near 10 K, while the Fermi-liquid $AT^2$ regime is attained at $T< 1.5$ K. The coefficient $A$ is enhanced by two orders of magnitude over that of a normal metal, which is a general rule in heavy-electron compounds. Measurements of the thermal and transport properties in a magnetic field [1,3] provide further evidence that the electron correlations are primarily of antiferromagnetic nature.

Inelastic neutron-scattering experiments have put the evidence for antiferromagnetic spin fluctuations on firm footing [4-6]. The fluctuation spectrum is quite complex as different energy scales are present. Spin-polarized neutron-scattering data on polycrystalline material [4] yield a quasi-elastic contribution centered at ~10 meV, which is related to the fluctuating local f-moment. The size of the fluctuating moment is of the order of 2 $\mu_B$/U-atom, which is not far from the value of the effective moment deduced from the high-temperature Curie-Weiss constant ($\mu_{eff}= 2.6\pm 0.2$ $\mu_B$/U-atom) [1]. Subsequent polarized and unpolarized neutron scattering measurements on single-crystalline samples [5] revealed a response centered at 5 meV, which is consistent with antiferromagnetic short-range order between nearest neighbor uranium atoms located in adjacent basal planes ($UPt_3$ has a hexagonal crystal structure). The antiferromagnetic correlations disappear above $T_{max}$, while in-plane ferromagnetic correlations persist till about 150 K. At yet a lower energy (0.2 meV) a second type of antiferromagnetic in-plane correlations was found at **Q**= (0.5,0,1) [6]. Surprisingly, at the same Q-vector, weak magnetic Bragg reflections were detected. This then provided evidence that, in pure $UPt_3$, small-moment antiferromagnetic order (SMAF) develops below a Néel temperature of ~6 K [6]. The size of the ordered moment is unusually small, $m= 0.02\pm 0.01$ $\mu_B$/U-atom. It is directed along the $a^*$-axis in the hexagonal basal plane. The magnetic unit cell consists of a doubling of the nuclear unit cell along the $a^*$-axis. More recently another type of correlations was observed near Q= 0 (forward direction) at low energies in a time-of-flight experiment [7]. These ferromagnetic correlations near Q= 0 have been interpreted in terms of the effect of low-lying fermion quasi-particles in the presence of strong spin orbit coupling.

Incipient magnetic order in UPt$_3$ was first detected by substitution studies [2]. By replacing Pt by isoelectronic Pd, pronounced phase-transition anomalies appear in the thermal and transport properties. Notably, the λ-like anomaly in $c(T)$ and the Cr-type anomaly in $\rho(T)$ give evidence for an antiferromagnetic phase transition of the spin-density-wave type. Neutron-diffraction experiments [8] carried out on a single-crystalline sample with optimal doping, U(Pt$_{0.95}$Pd$_{0.05}$)$_3$ ($T_{N,max}$= 5.8 K), confirmed the antiferromagnetic order. The ordered moment equals 0.6±0.2 μ$_B$/U-atom and is directed along the $a^*$-axis. By plotting the Néel temperatures, deduced from the $c(T)$ and $\rho(T)$ data, as functions of the Pd concentration, the border of the antiferromagnetic phase could be delineated [3]. Anomalies observed in the thermal and transport data restricted the antiferromagnetic order to the concentration range 2-7 at.% Pd. More recently, microscopic techniques, like neutron diffraction (this work) and μSR [9], have extended the lower Pd concentration limit to ~1 at % Pd. We have termed this magnetic order large-moment antiferromagnetic order (LMAF) in order to distinguish it from SMAF observed in pure UPt$_3$. The magnetic instability in UPt$_3$ can also be triggered by substituting Th for U [10-12]. Remarkably, the magnetic phase diagrams for the (U,Th)Pt$_3$ and U(Pt,Pd)$_3$ pseudobinaries are almost identical. This shows that the localization of the uranium moments is not governed by the unit-cell volume of these pseudobinaries (the unit-cell volume decreases upon Pd doping, while it increases upon alloying with Th). Long-range magnetic order also shows up when UPt$_3$ is doped with 5 at.% Au, while substituting 5 at.% Ir, Rh, Y, Ce or Os, does not induce magnetic order [13-15]. This indicates that a shape effect, i.e. the change in the c/a ratio, is the relevant control parameter for the occurrence of magnetic order.

The pronounced spin-fluctuation phenomena and the incipient magnetic order unambiguously demonstrate the proximity to a magnetic instability of UPt$_3$. Therefore, it came as a great surprise that the strongly-renormalized Fermi liquid is also unstable against superconductivity [16]. In the past decade many experiments have demonstrated that superconductivity in UPt$_3$ is unconventional [17]. The most important manifestations of unconventional superconductivity in UPt$_3$ are (i) the observation of power laws in the temperature variation of the superconducting properties, rather than the standard BCS exponential laws, (ii) the splitting of the superconducting transition in zero magnetic field, and (iii) the existence of three superconducting vortex phases in the magnetic field-temperature plane. In the past years, a number of phenomenological Ginzburg-Landau models have been worked out in order to understand the observed field and pressure variation of the three vortex phases [18]. The model which received the most attention is the so-called E-representation model, which is based on the coupling of a two-dimensional superconducting order parameter to a symmetry breaking field (SBF) [19]. The underlying mechanism is that a weak SBF lifts the degeneracy of the order parameter, which results in two superconducting phases in zero field. The key issue of the E-model is to identify the SBF and to prove that it couples to superconductivity. A natural candidate for the SBF is the SMAF, which was found to coexist with superconductivity [6]. Within the E-model, the splitting of the superconducting transition temperature $\Delta T_c = T_c^+ - T_c^-$ is proportional to the strength of the symmetry breaking field, $\Delta T_c \propto \varepsilon$, or in case that the SMAF acts as the SBF, $\Delta T_c \propto m^2$.

In this paper we report neutron-diffraction experiments conducted to investigate the evolution of magnetic order in the U(Pt,Pd)$_3$ series. The aim of these experiments was to answer the following questions: (i) what is the connection between the SMAF observed in pure UPt$_3$ and the LMAF observed in the doped compounds, (ii) how does the LMAF emerge upon Pd doping, (iii) is the SMAF stable with respect to Pd doping and does it couple to superconductivity, and (iv) is the SMAF influenced by annealing the samples. In order to address these questions we have carried out neutron diffraction experiments on single-crystalline U(Pt$_{1-x}$Pd$_x$)$_3$ with $x = 0.001, 0.002, 0.005, 0.01, 0.02$ and $0.05$. For all concentrations $x \leq 0.01$ we were able to detect SMAF, while for $x \geq 0.01$ LMAF was observed. The paper is organized as follows. In section 2 we focus on the experimental details, like the sample preparation process and the relevant information regarding the neutron scattering experiments. Section 3 is devoted to the calculation of the magnetic moment. In sections 4 and 5 our neutron diffraction results for the SMAF and the LMAF compounds are presented. In section 6 we constitute the magnetic phase diagram and in section 7 we discuss the connection between SMAF and superconductivity. In section 8 we discuss the results. A preliminary account of part of this work was presented in Ref. 20.

## 2. Experimental

Polycrystalline material was prepared by arc-melting the constituents in a stoichiometric ratio in an arc furnace on a water-cooled copper crucible under a continuously Ti-gettered argon atmosphere (0.5 bar). As starting materials we used natural uranium (JRC-EC, Geel) with a purity of 99.98%, and platinum and palladium (Johnson Matthey) with a purity of 99.999%. Polycrystalline material with low Pd contents ($x \leq 0.01$) was prepared by using appropriate master alloys (e.g. 5 at.% Pd). Single-crystalline samples with $x= 0.002, 0.01, 0.02$ and $0.05$, were pulled from the melt using a modified Czochralski technique in a tri-arc furnace under a continuously Ti-gettered argon atmosphere. Single-crystals with $x= 0.001$ and $0.005$ were prepared in a mirror furnace (NEC-NSC35) using the horizontal floating zone method. In order to anneal the samples, they were wrapped in tantalum foil and put in water free quartz tubes together with a piece of uranium that served as a getter. After evacuating ($p < 10^{-6}$ mbar) and sealing the tubes, the samples were annealed at 950 ºC during four days. Next the samples were slowly cooled in three days to room temperature. In the case of the samples with $x= 0.001$ and $x= 0.002$, neutron-diffraction data were collected before and after annealing. In all cases, the volume of the measured samples was of the order of 0.15 cm$^3$.

In order to characterize the samples the resistivity was measured on bar-shaped specimens spark-cut along the crystallographic a-and c-axis. The residual resistivity, $\rho_{0,a}$ and $\rho_{0,c}$, values are listed in Table I. For pure UPt$_3$ we obtain residual resistance ratios (RRR) of ~460 and ~720 for a current along the a-and c-axis, respectively. Upon alloying with Pd, $\rho_{0,a}$ increases smoothly with Pd content at a rate of 11.3 $\mu\Omega$cm/at.%Pd ($x \leq 0.01$), which shows that palladium is dissolved homogeneously in the matrix. Also the superconducting transition temperature ($T_c^+$) varies smoothly with Pd content, while the width $\Delta T_c^+$ stays about the same (see Table I). $T_c^+$ is suppressed at a rate 0.79 K/at.%Pd, and the critical concentration $x_c$ for the suppression of superconductivity equals 0.7 at.%Pd. Several crystals were investigated by Electron Probe Micro Analysis (EPMA), but the concentration of Pd is too small to arrive at a quantitative composition analysis. In the following sections the value of $x$ is taken as the nominal composition.

The neutron-diffraction experiments were carried out at three different reactor facilities. At Siloé (CEA-Grenoble) the samples with $x= 0.01, 0.02$ and $0.05$ were measured in the temperature range 1.8-10 K, using the DN1 triple-axis spectrometer. At the Institute Laue-Langevin in Grenoble the samples with $x= 0.002, 0.005$ and $0.01$ were measured in the temperature interval 0.1-10 K, using the IN14 triple-axis spectrometer. Finally, at the Laboratoire Léon Brillouin (CEA-Saclay) experiments were carried out on the samples with $x= 0.001$ and $0.002$ in the temperature range 0.1-10 K on the 4F2 triple-axis spectrometer.

For all experiments a pyrolytic graphite PG(002) analyzer was set to zero-energy transfer in order to separate the elastic Bragg scattering from possible low-energy magnetic excitations. To suppress the second order contamination a 10 cm long Be-filter and/or a 4 cm long pyrolytic graphite (PG) filter was used (see Table II). A vertically focusing PG(002) monochromator was used in all cases. The incident wave vector and the collimation of the different instruments are listed in Table II. The four different collimation angles refer to: (i) the collimation of the neutrons incident on the monochromator, (ii) collimation before the sample, (iii) collimation before the analyzer and (iv) collimation before the detector.

UPt$_3$ crystallizes in a hexagonal closed packed structure (MgCd$_3$-type) with space group P6$_3$/mmc [22]. The lattice parameters are given by a= 5.764 Å and c= 4.899 Å. The atomic positions in the unit cell are given by:

$$2 \text{ U at } \left(\frac{1}{3}, \frac{2}{3}, \frac{1}{4}\right)\left(\frac{2}{3}, \frac{1}{3}, \frac{3}{4}\right)$$

$$6 \text{ Pt at } \left(z, 2z, \frac{1}{4}\right)\left(2\bar{z}, \bar{z}, \frac{1}{4}\right)\left(z, \bar{z}, \frac{1}{4}\right) \quad (1),$$

$$\left(\bar{z}, 2\bar{z}, \frac{3}{4}\right)\left(2z, z, \frac{3}{4}\right)\left(\bar{z}, z, \frac{3}{4}\right)$$

where the ideal value of z equals 5/6. The Bragg positions are labeled using reciprocal lattice units, where a$^*$= b$^*$= 4π/(a√3) =1.264 Å$^{-1}$ and c$^*$= 2π/c = 1.283 Å$^{-1}$. In order to facilitate a quantitative analysis, the samples were always mounted with the c$^*$ axis vertical, i.e. perpendicular to the scattering plane. In the case of the samples with $x$= 0.005 and 0.01 additional data were taken with the reciprocal (1,-2,0) axis vertical, i.e. with the a$^*$-c$^*$ plane as the scattering plane.

## 3. Calculation of the magnetic moment

The neutron-diffraction experiments on pure UPt$_3$ [6] and the doped compounds U(Pt$_{0.95}$Pd$_{0.05}$)$_3$ [8] and (U$_{0.95}$Th$_{0.05}$)Pt$_3$ [12] show that the SMAF and LMAF have an identical magnetic structure. The magnetic unit cell corresponds to a doubling of the nuclear unit cell along the a$^*$-axis (with the moments pointing along the a$^*$-axis). This magnetic structure is schematically shown in Fig.1. In Fig.2 we have indicated the positions of the corresponding magnetic Bragg peaks in the reciprocal basal plane as observed by neutron scattering. The magnetic Bragg peaks corresponding to the domain with propagation vector **q$_1$**= (1/2,0,0) are located at e.g. **Q**= (1/2,1,0), (3/2,-1,0), (-1/2,-1,0) and (-3/2,1,0), as indicated by the open circles in Fig.2. Neutron scattering measures the projection of the Fourier component of the moment on a plane perpendicular to the scattering vector **Q**. For reflections such as (±1/2,0,0) this component **m**$_{q_1}$ is parallel to **Q** and the intensities vanish. There exist two other symmetry-related domains, **q$_2$** and **q$_3$**, obtained from **q$_1$** by a rotation of 120º and 240º, respectively. Assuming a single-q structure, **q$_1$**, **q$_2$** and **q$_3$** describe the three antiferromagnetic domains. In the absence of an in-plane magnetic field one expects, in general, to measure the same intensity for the magnetic Bragg peaks of the three domains. In this case the antiferromagnetic Fourier component, $\mathbf{m_q}$, becomes equal to the U magnetic moment, $\mathbf{m}$. We will comment on the possibility of a triple-q structure later.

A proper determination of the size of the (tiny) ordered magnetic moments across the $U(Pt_{1-x}Pd_x)_3$ series is not an easy task. Therefore, we have chosen to measure the various samples under the same experimental conditions and also to use the same procedure for the calibration of magnetic intensities. In order to determine the size of the magnetic moment, the cross sections of the magnetic and nuclear Bragg peaks have to be compared. We use the integrated intensity from longitudinal (θ-2θ) scans. The integrated nuclear $P_N$ and magnetic $P_M$ intensities are calculated from [23,24]:

$$P_N(\mathbf{Q}) = c\, L(\theta) \left| \sum_j b_j e^{(i\mathbf{Q}\cdot\mathbf{R}_j)} e^{-W_j} \right|^2 \tag{2a}$$

$$P_M(\mathbf{Q}) = c\, L(\theta) \left| p \sum_j \mathbf{m}^\perp_{\mathbf{q},j} f_j(\mathbf{Q}) e^{(i\mathbf{Q}\cdot\mathbf{R}_j)} e^{-W_j} \right|^2 \tag{2b}$$

where the sum is taken over all the Bravais lattices of the nuclear unit cell. $\mathbf{R}_j$ denotes the position of the nuclei in the cell, $L(\theta)=1/\sin(2\theta)$ is the Lorentz factor with θ the Bragg angle, $e^{-W_j}$ is the Debye-Waller factor, $b_j$ is the scattering length of the nucleus at site $j$, $f_j(\mathbf{Q})$ is the magnetic form factor, the symbol ⊥ denotes the projection on the plane perpendicular to the scattering vector $\mathbf{Q}$, $p= 0.2696\times10^{-12}$ cm, and $c$ is a normalization constant depending on the experimental conditions. For scattering in the basal plane there are two types of nuclear peaks which could be used for calibration, i.e. the (1,0,0) and (1,1,0)-type peaks. However, the intensity of the (1,0,0) reflection is very sensitive to deviations from the ideal Pt position $z= 5/6$ in the unit cell (see Fig.3). Actually, the measured ratio of the (1,0,0) and the (1,1,0) nuclear peaks indicates that the proper z-value is 0.8253 or 0.8370 instead of 5/6 (see Fig.3). We have chosen to use the (1,1,0) nuclear peak for calibration as its intensity depends only weakly on the z-value. By this procedure we possibly introduce a systematic error in determining the ordered moment. However this error is the same for all samples, so that a meaningful comparison between the moments of the samples can be made. The systematic error is not included in the error bars of the ordered moment for the different samples. Note that the variation of the lattice parameters a and c for $x \leq 0.05$ is almost negligible. The a parameter remains constant within the experimental accuracy and the c-parameter decreases at a relative rate of $0.7\times10^{-4}$ per at.% Pd [3].

# 4. Small-moment antiferromagnetic order for $0 \leq x \leq 0.01$

Neutron-diffraction experiments have been carried out in the temperature range 0.1-10 K on annealed U(Pt$_{1-x}$Pd$_x$)$_3$ single-crystals with $x=$ 0.005 and 0.01 and unannealed crystals with $x=$ 0.001 and 0.002. The samples with $x=$ 0.001 and 0.002 were remeasured in the temperature interval 1.8-10 K after annealing. In Fig.4 we have plotted the temperature variation of the maximum intensity of the magnetic Bragg peak at **Q**= (1/2,1,0) for $x \leq 0.01$ after subtracting the background. Let us first focus on the data of the annealed samples, represented by open symbols. In this case, absolute values of $m^2$ in units of $\mu_B^2$ have been plotted using the calibration procedure presented in section 3.

The behavior of $m^2(T)$ for the various samples as shown in Fig.4 is quite unusual. Fig.4 clearly demonstrates that the small-moment magnetism is robust upon alloying with Pd. The size of the ordered moment increases gradually with Pd concentration, but, remarkably, SMAF invariably sets in near $T_N \sim$ 6 K for $x \leq 0.01$. For all samples with $x \leq 0.005$, $m^2(T)$ has an unusual form. The value of $m^2$ starts to rise slowly below $T_N \sim$ 6 K, then a quasi-linear temperature dependence follows from ~4 K down to $T_c^+$ (0.1-0.4 K, see Table I). Below $T_c^+$ the magnetic intensity saturates. The absolute values of the ordered moments have been calculated using integrated intensities. We obtain $m(T_c^+)=$ 0.024±0.003, 0.036±0.003 and 0.048±0.008 $\mu_B$/U-atom for $x=$ 0.001, 0.002 and 0.005, respectively, in the annealed state (see also Table I). For comparison Fig.4 shows also $m^2(T)$ for pure UPt$_3$, as reported by Hayden et al. [25]. The value for $m(T_c^+)$ was estimated in Ref. 25 at 0.03±0.01 $\mu_B$/U-atom. Because of the relatively large uncertainty in this value we have calibrated $m^2(T)$ for pure UPt$_3$ with help of a recent measurement by Van Dijk et al. (Ref. 26). Following the same calibration procedure as for the doped compounds we arrive at the value $m=$ 0.018±0.002 $\mu_B$/U-atom for pure UPt$_3$. This is identical to the value reported recently by Isaacs et al. (Ref. 27).

The effect on annealing was investigated for the $x=$ 0.001 and 0.002 samples. In the limit $T \rightarrow T_c^+$ $m$ equals 0.019±0.003 and 0.038±0.003 $\mu_B$/U-atom in the unannealed state, for $x=$ 0.001 and 0.002, respectively. This shows that the size of the ordered moment is not changed (within the experimental accuracy) by our annealing procedure. Also the temperature variation of $m^2(T)$ does not change upon annealing. This is illustrated by the comparison of the data for the annealed and unannealed samples shown in Fig.4, where the moments for the unannealed sample have been multiplied by a factor 1.26 and 0.95, for $x=$ 0.001 and 0.002, respectively, for normalization purposes (assuming that in the limit $T \rightarrow T_c^+$ $m$ is the same for annealed and unannealed samples).

In order to investigate the effect of annealing on the magnetic correlation length, $\xi_m$, we have scanned the magnetic Bragg peak at **Q**= (1/2,1,0) at several selected temperatures for $x$= 0.001 and 0.002 before and after annealing. Typical data sets, taken on the annealed $x$= 0.001 and 0.002 samples, are shown in Fig.5 and Fig.6, respectively. By fitting a Lorentzian profile, convoluted with the Gaussian experimental resolution, we were able to extract the correlation length along **Q**. Note that the width of the $\lambda/2$ peak, measured without the Be filter, is not a correct estimate for the experimental resolution on the spectrometers used here (see Fig.6). For $x$= 0.001 we obtain $\xi_m$= 570±130 Å and $\xi_m$= 710±150 Å before and after annealing, and for $x$= 0.002 $\xi_m$= 700±150 Å and $\xi_m$= 610±130 Å before and after annealing. Thus no effect of annealing on $\xi_m$ is observed within the experimental error. This is consistent with the recent conclusion reached by Isaacs et al. [27], who investigated the effect of annealing on the correlation lengths along $a^*$ and $c^*$ for pure UPt$_3$. Since the size of the ordered moments and the values of the correlation lengths are within the experimental error the same before and after annealing, we conclude that strain has no significant effect on the SMAF.

## 5. Large-moment antiferromagnetic order for $x \geq 0.01$

In this section we report our neutron-diffraction results on the annealed U(Pt$_{1-x}$Pd$_x$)$_3$ single crystals with $x$= 0.01, 0.02 and 0.05. We have plotted the temperature variation of the maximum intensity of the magnetic Bragg peak at **Q**= (1/2,1,0) (background subtracted) for $x$= 0.02 and 0.05 in Fig.7 and for $x$= 0.01 at **Q**= (1/2,0,1) in Fig.8. Absolute values of $m^2$ in units of $\mu_B^2$ have been obtained using the calibration procedure presented in section 3. The temperature variation $m^2(T)$ for $x$= 0.02 and 0.05 is rather conventional compared to the quasi-linear temperature variation observed for the SMAF compounds (Fig.4). The order parameter follows $m^2(T) \propto (1-(T/T_N)^\alpha)^{2\beta}$, with the values $\alpha$= 1.9±0.2 and 1.8±0.1 and $\beta$= 0.50±0.05 and 0.32±0.03 for $x$= 0.02 and 0.05, respectively. These values of $\beta$ are not too far from the theoretical value $\beta$= 0.38 for the 3D Heisenberg model [28]. The phenomenological parameter $\alpha$ reflects spin-wave excitations. In a cubic antiferromagnetic system $\alpha$ is predicted to be 2 [29]. To our knowledge no predictions are available for a hexagonal system. In the limit $T \rightarrow 0$ K, we obtain $m$= 0.35±0.05 and 0.63±0.05 $\mu_B$/U-atom for $x$= 0.02 and 0.05, respectively. The size of the ordered moment obtained for $x$= 0.05 is in excellent agreement with the value reported in Ref. 8. For the LMAF compounds the magnetic Bragg peaks are resolution-limited.

The temperature dependence of the magnetic Bragg intensity of the sample with $x$= 0.01 is quite intriguing: $m^2(T)$ starts to rise slowly below $T_N$~ 6 K, grows rapidly below ~2 K, and then saturates below ~0.5 K. The rapid rise near 2 K suggests a cross-over from the small-moment to the large-moment state, with an estimate of $T_N$~ 1.8 K for the LMAF. For $T \to 0$ K, $m$ reaches a value of 0.11±0.03 $\mu_B$/U-atom. This value is obtained for both $\mathbf{Q}$= (1/2,1,0) and $\mathbf{Q}$= (1/2,0,1). We emphasize that the width of the magnetic Bragg peak does not change in the temperature range 0.08-3 K (see Fig.9), which ensures that the unusual $m^2$(T) curve is not due to an increase of $\xi_m$ upon lowering the temperature. The interpretation of a cross-over to the LMAF state is consistent with recent µSR experiments on U(Pt$_{0.99}$Pd$_{0.01}$)$_3$ [9], which show that the LMAF gives rise to a spontaneous $\mu^+$ precession frequency below $T_N$~ 1.8 K.

In the case of $x$= 0.01, the transition to the LMAF state does not show up in the thermal and transport data, in contrast to data for $x$= 0.02 and 0.05, which exhibit clear magnetic phase transitions at $T_N$= 3.5 and 5.8 K, respectively [2,3]. Careful resistivity measurements down to 0.016 K on a polycrystalline sample with composition U(Pt$_{0.99}$Pd$_{0.01}$)$_3$ did not reveal any signature of a phase transition [30]. This was taken as evidence that the Néel temperature for the LMAF drops to zero between 1 and 2 at.% Pd. However, the present neutron-diffraction data show that the lower bound for LMAF is actually between 0.5 and 1 at.% Pd.

# 6. Evolution of magnetism in the U(Pt$_{1-x}$Pd$_x$)$_3$ pseudobinaries

Our neutron-diffraction results show that all the U(Pt$_{1-x}$Pd$_x$)$_3$ compounds ($x \leq$ 0.05) order antiferromagnetically. In Fig.10 we plot the Néel temperatures of the different samples versus Pd concentration. For samples with $x \leq$ 0.01 SMAF invariably sets in with a Néel point of ~6 K. Most likely this phase line extends horizontally to higher Pd concentrations, but for $x$> 0.01 it becomes more and more difficult to discriminate experimentally between SMAF and LMAF. A closer inspection of the data for $x$= 0.02 (Fig.7) shows that indeed some magnetic intensity is visible in the temperature range 3.5-6 K. However, a careful measurement of the background signal for $x$= 0.02 is needed in order to put this on firm footing. LMAF emerges in the concentration range 0.5-1 at.% Pd. The optimum doping for LMAF is 5 at.% Pd. This compound has the largest Néel temperature, $T_N$= 5.8 K, and magnetic moment, $m$= 0.63±0.05 $\mu_B$/U-atom. For $x$= 0.10 no LMAF has been observed in the thermal and transport properties. However, at this moment, we cannot exclude LMAF with a reduced $T_N$ as observed for $x$= 0.01. In order to investigate the Pd rich side of the phase diagram, neutron-diffraction or µSR experiments would be most welcome. On the other hand, one should keep in mind that additional lines in the x-ray diffraction patterns indicate that the MgCd$_3$-type of structure is lost for $x \geq$ 0.15 [3]. $T_N$ for the LMAF follows a rather conventional Doniach-type phase diagram [33]. The compound with $x$= 0.01 occupies a special place in the phase diagram as we have assigned two Néel temperatures to it. The SMAF which emerges near 6 K develops into LMAF near 1.8 K.

The size of the ordered moment, measured at $T_c^+$ as function of Pd concentration is plotted in Fig.11. The moment first increases slowly from 0.018±0.002 $\mu_B$/U-atom for pure UPt$_3$ to 0.036±0.003 $\mu_B$/U-atom for 0.5 at.% Pd. For higher Pd concentrations the moment rises much more rapidly. The change in slope of $m(x)$ between $x=0.005$ and $x=0.01$ is consistent with LMAF emerging in this concentration range.

## 7. Interplay of magnetism and superconductivity

Recently, we have measured the specific heat and electrical resistivity at the superconducting transition of single-crystalline ($x=0.0, 0.001, 0.002$ and $0.005$) and polycrystalline ($x=0.0025, 0.003, 0.004, 0.006$ and $0.007$) UPt$_3$ doped with small amounts of Pd [34,35]. The main findings can be summarized by (i) $T_c^+$ is suppressed linearly with Pd content at a rate of 0.79±0.04 K/at.%Pd, (ii) $T_c^-$ is suppressed at a faster rate of 1.08±0.06 K/at.%Pd, and as a results (iii) the splitting $\Delta T_c$ *increases* at a rate 0.30±0.02 K/at.%Pd. This shows that upon alloying with Pd, the high-temperature low-field A phase gains stability at the expense of the low-temperature low-field B phase. The data in Fig.4 show that the increase in $\Delta T_c$ is accompanied by an increase in the size of the ordered moment. This provides additional support to the idea that the SMAF acts as the symmetry breaking field. The Ginzburg-Landau E-representation scenario [19] predicts $\Delta T_c \propto m^2$. However, this proportionality relation is only valid for $\Delta T_c/T_c \ll 1$, which no longer holds for the Pd-doped samples. At ~0.3 at.% Pd, $\Delta T_c$ becomes of the order of $T_c$. Instead $m^2$ grows more rapidly than $\Delta T_c$. Substantial evidence for the SMAF as the symmetry breaking field has been obtained by neutron-diffraction [25] and specific-heat [36] experiments under pressure. It was found that both $m^2$ and $\Delta T_c$ are suppressed quasi-linearly with pressure and vanish at a critical pressure p$_c$~ 0.35 GPa. Interestingly, we find a smooth variation of $\Delta T_c$ as function of $m^2$ when we collect both the pressure and Pd doping data [35]. This establishes a firm link between $\Delta T_c$ and $m^2$. Only for small splittings is $\Delta T_c \propto m^2$ ($\Delta T_c < 0.050$ K). For enhanced splittings a more sophisticated Ginzburg-Landau expansion (with terms beyond 4th. order) should be elaborated.

The critical Pd concentration $x_c$ for the suppression of superconductivity is ~0.7 at.% Pd [35]. The value of $x_c$ falls in the range where LMAF emerges. It would be of interest to know whether the suppression of superconductivity coincides with the emergence of LMAF. µSR experiments aimed at probing the LMAF in this concentration range are in progress.

## 8. Discussion

Our neutron-diffraction data unambiguously show that the unusual small-moment antiferromagnetic order observed in pure UPt$_3$ is stable upon Pd doping. Indeed, we find that Pd doping leads to an enhancement of SMAF as the ordered moment grows with increasing Pd content. The reverse behavior was observed in the neutron-diffraction experiments under pressure carried out on pure UPt$_3$ [25]. The moment decreases under pressure and vanishes completely at $p_c \sim 0.35$ GPa. A quite remarkable observation is that both data sets, obtained by Pd doping and applying pressure, show that $T_N$ retains a constant value of ~6 K. This, together with the gradual increase of $m^2(T)$ below ~6 K, could indicate that the transition to the SMAF state is not a true phase transition.

The origin and nature of the SMAF are still subjects of lively debates. Unraveling the nature of the small moment is hampered by the fact that, till today, it has been probed convincingly by neutron-diffraction (Refs. 6, 25-27 and this work) and magnetic x-ray scattering [27] experiments only. The analysis of both neutron-diffraction and magnetic x-ray scattering data [27], lead to the conclusion that the SMAF is quantitatively the same in the bulk and near surface of annealed UPt$_3$. The only difference is the somewhat smaller correlation length along a$^*$ and c$^*$ obtained in the case of magnetic x-ray scattering, $\xi_{a^*} = 85 \pm 13$ Å and $\xi_{c^*} = 113 \pm 30$ Å at $T = 0.15$ K. These values should be compared to $\xi_{a^*} = 280 \pm 50$ Å and $\xi_{c^*} = 500 \pm 130$ Å at $T = 0.57$ K in the case of the neutron diffraction experiment.

The possibility that the small moment is caused by magnetic impurities, defects or sample inhomogeneities can safely be excluded. Firstly, rather high impurity concentrations would be needed, for instance, ~1000 ppm of magnetic impurities with moments of 0.6 $\mu_B$, in order to obtain the same magnetic signal as for the small moment of 0.02 $\mu_B$. Secondly, impurities will not contribute to Bragg peaks of the type (1/2,0,0), since randomly distributed impurities or defects would give Q-independent scattering, while ordered imperfections would give rise to new satellite Bragg peaks close to the nuclear peaks. The same arguments are valid for stacking faults, observed in polycrystalline materials by transmission electron microscopy and x-ray diffraction measurements [37], and which could locally change the crystal symmetry and give rise to magnetic moments on certain uranium atoms. On the other hand, one can imagine that there are sizable sample regions (clusters) where large magnetic moments develop, which are perfectly ordered with a propagation vector of (1/2,0,0). This in principle could give rise to the observed Bragg peaks. Due to the finite size of these clusters (100-500 Å), the magnetic correlation length is limited. These clusters would form 0.1% of the sample volume and would be separated by large regions of non-magnetic UPt$_3$. However, the minor influence of annealing on the SMAF, and the fact that the better samples (as determined by the degree of crystallographic order) all exhibit a magnetic moment [38], strongly suggest that SMAF is an intrinsic property.

At this point it is important to note that recent zero-field μSR studies on polycrystalline [8] and single-crystalline [39] UPt$_3$ failed to detect the small magnetic moment (except for the μSR study reported in Ref. 40, but this result has not been reproduced). In the course of a detailed investigation [9] of the evolution of magnetism in U(Pt,Pd)$_3$ by the μSR technique, we found that LMAF gives rise to a spontaneous μ$^+$ precession frequency. However, we did not observe any signal of the SMAF in polycrystalline samples with $x$=0.000, 0.002 and 0.005. A possible explanation for this is that the muon comes to rest at a site where the magnetic dipolar fields cancel due to the magnetic ordering. However, this is highly unlikely as SMAF and LMAF have an identical magnetic structure and we have been able to detect the LMAF (in samples with $x$= 0.01, 0.02 and 0.05). It is also unrealistic to expect a change of the stopping site at these low Pd concentrations. The μSR technique is sensitive enough to detect a static moment of the order of 0.02 μ$_B$. One possibility is that the small moment fluctuates at a rate ($f$ > 10 MHz) too fast to be detected by μSR, but on a time scale which appears static to neutrons and x-rays. This then also solves the long-standing problem of why the small moment of UPt$_3$ cannot be seen by NMR, while its signal should fall well in the detection limit as was concluded from experiments on U(Pt$_{1-x}$Pd$_x$)$_3$ ($x$≤ 0.05) which successfully probed the LMAF [41]. Fluctuating moments are also in line with the hypothesis that there is no true phase transition at $T_N$ for SMAF. The invariance of $T_N$ and the cross-over-type of behavior suggests that the small moment is only a weak instability of the renormalized Fermi-liquid whose properties hardly change at these low Pd concentrations ($x$≤ 0.005).

In the Ginzburg-Landau analysis [19], which makes use of the symmetry breaking field scenario, it is generally assumed that the SMAF forms in a single-q structure. However, the existing neutron scattering data are compatible with a triple-q structure as well. The question whether the magnetic order corresponds to a single-q or a triple-q structure is crucial for the understanding of the unconventional superconductivity because a single-q structure breaks the hexagonal symmetry, while a triple-q does not. The single-q and triple-q structures can be distinguished by applying a magnetic field. For example, in the case of a strong magnetic field applied along the b-axis, one expects to re-orient all domains along the a$^*$-axis or in the terms of Fig.2, **q$_1$** is expected to increase a factor 3 due to the depopulation of **q$_2$** and **q$_3$**. Experiments carried out up to 3.2 T [42] and 12 T [26] did not show any redistribution of magnetic domains, so a triple-q structure for the SMAF cannot be excluded. However, it is possible that a field of 12 T is not sufficiently strong to change the domain population of moments as weak as 0.02 $\mu_B$. The SMAF itself is very stable to a magnetic field. $T_N$ is suppressed by only 0.7 K and 0.4 K for a field of 10 T applied along the a and c-axis, respectively. In the case of the LMAF the magnetic structure is single-q. Neutron-diffraction experiments [43] carried out on U(Pt$_{0.95}$Pd$_{0.05}$)$_3$ as function of an external field applied in the basal plane showed the formation of a single-domain sample in 5 T.

The magnetic phase diagram of the U(Pt$_{1-x}$Pd$_x$)$_3$ pseudobinaries (Fig.10) is quite unusual because of the distinction between SMAF and LMAF. The differences between the SMAF and LMAF can be outlined as follows: (i) the order parameter for the SMAF is unusual and grows quasi-linearly, while the order parameter for the LMAF is conventional and confirms a real phase transition, (ii) $T_N$ for the SMAF does not change with Pd content, while $T_N$ of the LMAF compounds follows a rather conventional Doniach-type phase diagram, (iii) the SMAF is not observed in zero-field µSR experiments in contrast to the LMAF. This demonstrates that the SMAF and LMAF are not directly connected.

While the origin of SMAF in UPt$_3$ remains unclear, the emergence of LMAF in the alloyed systems is a general feature of heavy-fermion systems. The magnetic instability is normally explained in terms of a competition between the on-site Kondo effect and the inter-site Ruderman-Kittel-Kasuya-Yosida (RKKY) interaction. However, in the case of the U(Pt,Pd)$_3$ system a clear-cut identification of $T_K$ and $T_{RKKY}$ is not at hand [44]. Moreover, since UPt$_3$ is very close to a magnetic instability, the variation of $T_K$ and $T_{RKKY}$ before magnetic ordering occurs is small. Better documented systems in this respect are (Ce$_{1-x}$La$_x$)Ru$_2$Si$_2$, where magnetism sets in at $x= 0.07$ [45] and CeCu$_{6-x}$Au$_x$, where magnetism sets in at $x= 0.1$ [46]. In these systems the magnetic instability is reached at a critical hybridization, which results from expanding the lattice. In the case of U(Pt,Pd)$_3$ the occurrence of LMAF can be parametrized, to a certain extent, by the reduction of the c/a ratio upon alloying (and not by a volume effect, as the volume decreases). The application of pressure has the opposite effect, since pressure increases the c/a ratio due to the anisotropy in the linear compressibilities ($\kappa_c < \kappa_a$) [3]. These effects are however small and a satisfactory quantitative analysis is hampered by the limited accuracy in the values of the lattice constants and compressibilities. Pressure experiments, carried out on the 5 and 7 at.% Pd samples show that doping 1 at.% Pd corresponds to an external pressure of about -0.33 GPa [47]. In the case of 5 at.% Pd it was demonstrated by specific-heat experiments under pressure [48] that the LMAF state was fully suppressed at ~1.6 GPa, thereby recovering the situation of pure UPt$_3$.

Currently, much attention in heavy-fermion research is focused on the occurrence of non-Fermi-liquid effects at the critical concentration for the suppression of magnetism. In the case of U(Pt,Pd)$_3$ we expect that the border line magnetic/non-magnetic is close to 0.7 at.% Pd, which is also the critical concentration for the suppression of superconductivity. Resistivity and specific-heat experiments performed so far did not show any signature of non-Fermi-liquid phenomena. However, the quantum critical point has not been probed in full detail yet.

## 9. Summary

Neutron-diffraction experiments, carried out on a series of heavy-electron pseudobinary U(Pt$_{1-x}$Pd$_x$)$_3$ single crystals ($x \leq 0.05$), show that two kinds of antiferromagnetic order, termed small-moment antiferromagnetic order (SMAF) and large-moment antiferromagnetic order (LMAF), are found in the phase diagram. The small-moment antiferromagnetic order, first reported for pure UPt$_3$, is robust upon doping with Pd and persists till at least $x= 0.005$. The ordered moment grows from $0.018 \pm 0.002$ $\mu_B$/U-atom for pure UPt$_3$ to $0.048 \pm 0.008$ $\mu_B$/U-atom for $x= 0.005$. The Néel temperature of 6 K, does not vary with Pd contents. The order parameter for the small-moment antiferromagnetism has an unusual quasi-linear temperature variation and points to a cross-over phenomenon rather than a true phase transition. The small moment is not observed by µSR and NMR experiments. This could indicate that the moment is not static, but fluctuates at a rate larger than 10 MHz. For $x \geq 0.01$ large-moment antiferromagnetic order is observed. At the optimum doping ($x= 0.05$) $T_N$ attains a maximum value of 5.8 K and the ordered moment equals $0.63 \pm 0.05$ $\mu_B$/U-atom. $T_N(x)$ for the large-moment antiferromagnetic order follows a Doniach-type phase diagram. From this diagram we infer that the antiferromagnetic instability in U(Pt$_{1-x}$Pd$_x$)$_3$ takes place for Pd concentrations $0.005 < x < 0.01$.

**Acknowledgments**

This work was part of the research program of the Dutch Foundation for Fundamental Research of Matter ("Stichting" FOM). J. Bossy, S. Pujol, N.H. van Dijk and Ph. Boutrouille are gratefully acknowledged for experimental assistance at various stages of these experiments. We thank J. Flouquet and P.H. Frings for stimulating discussions. R.J. Keizer and A. de Visser acknowledge the EC-TMR Large Scale Facilities program at LLB for financial support.

Table I  Some characteristic properties of the annealed single-crystalline U(Pt$_{1-x}$Pd$_x$)$_3$ samples. The residual resistivity, $\rho_{0,a}$ and $\rho_{0,c}$, the upper superconducting transition temperature $T_c^+$, and its width $\Delta T_c^+$ as determined by transport experiments [21], the superconducting splitting, $\Delta T_c = T_c^+ - T_c^-$, determined by the specific heat, and the magnetic moment $m$ at $T_c^+$.

| $x$ | $\rho_{0,a}$ (μΩcm) | $\rho_{0,c}$ (μΩcm) | $T_c^+$ (K) | $\Delta T_c^+$ (K) c-axis | $\Delta T_c$ (K) | $m(T_c^+)$ (μ$_B$/U-atom) |
|---|---|---|---|---|---|---|
| 0.000 | 0.52(5) | 0.18(3) | 0.543 | 0.006(1) | 0.054(4) | 0.018(2) |
| 0.001 | 1.6(2) | 0.75(6) | 0.437 | 0.009(1) | 0.082(4) | 0.024(3) |
| 0.002 | 2.5(2) | 1.02(9) | 0.384 | 0.007(1) | 0.108(5) | 0.036(3) |
| 0.005 | 6.2(5) | - | 0.126 | - | - | 0.048(8) |

Table II  Specifications of the spectrometers used in the experiments.

| Facility | k$_i$ (Å$^{-1}$) | collimation | filters |
|---|---|---|---|
| Siloé | 2.66 | open-30'-60'-60' | PG |
| ILL | 1.48 | 34'-40'-40'-40' | Be & PG |
| LLB | 1.48 | open-open-60'-60' | Be & PG |

**Figure captions**

Fig.1  Magnetic structure of U(Pt$_{1-x}$Pd$_x$)$_3$. The open and closed circles indicate U atoms in adjacent hexagonal planes separated by a lattice spacing c/2. The arrows indicate the magnetic moments, which are directed along the a$^*$-axis. The dotted and solid line delineate the nuclear and magnetic unit cell, respectively.

Fig.2  Reciprocal lattice (a$^*$-b$^*$ plane) of U(Pt$_{1-x}$Pd$_x$)$_3$. The open symbols indicate the positions where magnetic Bragg reflections are observed by neutron scattering. The three magnetic domains (assuming a single-q structure) are indicated by **q$_1$** (O), **q$_2$** (□) and **q$_3$** (Δ). The closed symbols indicate the positions of the nuclear (1,0,0) (■) and (1,1,0)-type (●) of reflections.

Fig.3  Calculated intensities of the nuclear (1,0,0) (solid line) and (1,1,0) (dashed line) Bragg peaks as function of the position, z, of the Pt atoms in the unit cell. From the measured ratio of the intensities for the (1,0,0) and the (1,1,0) Bragg peaks we find z= 0.8253 or z= 0.8370, instead of the ideal value z= 5/6 (indicated by the dotted vertical lines).

Fig.4  Temperature variation of $m^2$ derived from the intensity of the magnetic Bragg peak for annealed (open symbols) and unannealed (closed symbols) U(Pt$_{1-x}$Pd$_x$)$_3$. For $x$=0.001 (O), 0.002 (□), 0.005 (Δ) data are taken at **Q**= (1/2,1,0) and for $x$= 0.01 (◊) at **Q**= (1/2,0,1). In the case of $x$= 0.00 we have reproduced the data of Ref. 25 (dashed line) after normalizing them to the moment deduced in Ref. 26 (∇). The solid lines are guides to the eye.

Fig.5  Longitudinal scans of the magnetic Bragg peak **Q**= (1/2,1,0) for annealed U(Pt$_{0.999}$Pd$_{0.001}$)$_3$ at temperatures 1.6≤ $T$≤ 6.2 K as indicated. The solid lines are fits to the data using a Lorentzian convoluted with the Gaussian experimental resolution. The width of the λ/2 peak without Be filter is shown in the lower part of the figure together with the experimental resolution (dashed line).

Fig.6   Longitudinal scans of the magnetic Bragg peak $\mathbf{Q}=(1/2,1,0)$ for annealed U(Pt$_{0.998}$Pd$_{0.002}$)$_3$ at temperatures $1.7 \leq T \leq 5.3$ K as indicated. The solid lines are fits to the data using a Lorentzian convoluted with the Gaussian experimental resolution.

Fig.7   Temperature variation of $m^2$ for annealed U(Pt$_{1-x}$Pd$_x$)$_3$ derived from the intensity of the magnetic Bragg peak $\mathbf{Q}=(1/2,1,0)$ for $x=$ 0.02 ($\square$) and 0.05 ($\bigcirc$) and at $\mathbf{Q}=(1/2,0,1)$ for 0.01 ($\lozenge$). The solid lines represent fits to $m^2(T) \propto (1-(T/T_N)^\alpha)^{2\beta}$ (see text).

Fig.8   Temperature variation of $m^2$ measured at the magnetic Bragg peak $\mathbf{Q}=(1/2,0,1)$ for annealed U(Pt$_{1-x}$Pd$_x$)$_3$ with $x=0.01$ ($\lozenge$). The sharp increase in the intensity near 1.8 K indicates a crossover from SMAF to LMAF.

Fig.9   Longitudinal scans of the magnetic Bragg peak $\mathbf{Q}=(1/2,0,1)$ for annealed U(Pt$_{0.99}$Pd$_{0.01}$)$_3$ at temperatures $0.08 \leq T \leq 3$ K as indicated. The solid lines are fits to the data using a Lorentzian convoluted with the Gaussian experimental resolution. The horizontal arrows show the total width (FWHM) of the peak.

Fig.10  The Néel temperature, $T_N$, versus Pd concentration for U(Pt$_{1-x}$Pd$_x$)$_3$ alloys as determined from neutron diffraction ($\bigcirc$) and specific heat experiments ($\square$) (Refs. 2, 3, 31, 32). SMAF and LMAF denote small-moment and large-moment antiferromagnetic order, respectively. In the lower left corner the upper superconducting transition temperature $T_c^+$ as determined by resistivity experiments is given [35]. SC denotes the superconducting phase.

Fig.11  Uranium ordered moment at $T_c^+$ as function of Pd concentration in single-crystalline U(Pt$_{1-x}$Pd$_x$)$_3$ alloys. The line is a guide to the eye.